\documentclass[preprint,onecolumn,nofootinbib]{revtex4}
\usepackage[colorlinks=true,linkcolor=blue,urlcolor=blue,filecolor=black,citecolor=red,pdfstartview=FitV,pdftitle={},pdfsubject={},pdfkeywords={},pdfpagemode=None,bookmarksopen=true]{hyperref}
\usepackage{graphicx}
\usepackage{amsmath}
\usepackage{amsfonts}
\usepackage{amssymb,ulem}
\usepackage{color,xcolor}%
\usepackage{CJK}
\usepackage{subfigure}
\usepackage{amsthm,amsmath,amssymb}
\usepackage{mathrsfs}
\usepackage{multirow}
\usepackage{dcolumn}
\usepackage{float}
\newcommand{\blue}{\textcolor[rgb]{0.33,0.33,1.00}}

\begin{document}
\title{Probing Quantum Gravity Effects with Eccentric Extreme Mass-Ratio Inspirals}

\author{Guoyang Fu $^{1}$}
\thanks{fuguoyangedu@sjtu.edu.cn}
\author{Yunqi Liu $^{2}$}
\thanks{yunqiliu@yzu.edu.cn}
\author{Bin Wang $^{2,3}$}
\thanks{wang$\_$b@sjtu.edu.cn}
\author{Jian-Pin Wu $^{2}$}
\thanks{jianpinwu@yzu.edu.cn}
\author{Chao Zhang $^{4}$}
\thanks{zhangchao666@sjtu.edu.cn}
\affiliation{$^1$~ School of Physics and Astronomy, Shanghai Jiao Tong University,  Shanghai 200240, China}
\affiliation{$^2$~Center for Gravitation and Cosmology, College of Physical Science and Technology, Yangzhou University, Yangzhou 225009, China}
\affiliation{$^3$~Shanghai Frontier Science Center for Gravitational Wave Detection, Shanghai Jiao Tong University, Shanghai 200240, China}
\affiliation{$^4$~Department of Physics, School of Physical Science and Technology, Ningbo University, Ningbo, Zhejiang 315211, China}

\begin{abstract}

In this paper, we investigate the impact of loop quantum gravity (LQG) on extreme mass-ratio inspirals (EMRIs), and the results indicate that LQG effects cause the orbital decay to occur faster compared to the Schwarzschild case.
Furthermore, we use the augmented analytic kludge approach to generate EMRI waveforms and study the LISA's capability to detect the LQG effect with faithfulness.
Additionally, employing the Fisher information matrix method for parameter estimation, we estimate that after one-year observation, the uncertainty in $r_0$ reduces to approximately $6.59\times 10^{-4}$ with a signal-to-noise ratio of $49$.

\end{abstract}

\maketitle

\section{Introduction}\label{sec-intro}

General relativity (GR) is renowned for its revolutionary achievement in redefining gravity as the curvature of spacetime, fundamentally altering our comprehension of this fundamental interatcion. Two of the most remarkable predictions of GR, both confirmed through observations, are the existence of black holes (BHs) and the phenomenon of gravitational waves (GWs). Notably, GWs from binary system mergers have been detected \cite{LIGOScientific:2016aoc,LIGOScientific:2016lio,LIGOScientific:2016sjg}, and the Event Horizon Telescope has imaged the shadows of supermassive black holes (SMBH) M87$^*$ and $\mathrm{Sgr\ A^{*}}$ \cite{EventHorizonTelescope:2019dse,EventHorizonTelescope:2019ths,EventHorizonTelescope:2022xnr,EventHorizonTelescope:2022xqj}, thereby solidifying GR's status as a cornerstone of modern physics.

Extreme mass-ratio inspirals (EMRIs) are regarded as one of the most promising sources for future space-based GW detectors such as the Laser Interferometer Space Antenna (LISA) \cite{Danzmann:1997hm,LISA:2017pwj}, TianQin \cite{Luo:2015ght}, and Taiji \cite{Hu:2017mde}.
Without a doubt, the GWs radiated by EMRIs carry information about the spacetime geometry. Utilizing this information will offer unprecedented insights into a broader range of GW sources, enabling the detection of deviations from BH predictions, probing fundamental physics, and even potentially revealing quantum gravity effects.
\cite{Gong:2021gvw,TianQin:2020hid,Ruan:2018tsw,LISA:2022yao,LISA:2022kgy,Ghosh:2024arw, AbhishekChowdhuri:2023rfv, Rahman:2023sof, Rahman:2021eay, Rahman:2022fay, AbhishekChowdhuri:2023gvu, Zhang:2024csc, Zhang:2024ogc,Kumar:2024utz,Speri:2024qak}.  Refs. \cite{Maselli:2020zgv,Maselli:2021men,Guo:2022euk,Zhang:2022rfr,Brito:2023pyl,Liang:2022gdk,Zhang:2023vok,Zi:2024mbd} suggests that EMRIs can effectively detect whether small compact object carry scalar or vector charges. Additionally, EMRIs can also be employed to test and constrain modified gravity. For more details, we can refer to  \cite{Yunes:2011aa,Guo:2023mhq,Qiao:2024gfb,Berti:2005qd,Sopuerta:2009iy,Pani:2011xj,DeFalco:2023djo} and references therein.

Theoretically, the singularity theorem \cite{Hawking:1966sx,Penrose:1964wq} in GR highlights the inevitable emergence of spacetime singularities, signaling the breakdown of GR in these extreme conditions. Consequently, it is reasonable to expect that a quantum theory of gravity, which unifies quantum mechanics with GR, will be essential in these regimes and could potentially resolve the singularity problem.
Loop quantum gravity (LQG) is particularly distinguished for its background independence and non-perturbative nature, making it a promising and prominent framework in the field. The quantization techniques developed within the full LQG framework have been
successfully applied to the spherically symmetric Schwarzschild BH (SS-BH) model, signifying the inception of a novel research domain---loop quantum gravity black holes (LQG-BHs). For a detailed construction of the LQG-BHs, see \cite{Ashtekar:2005qt,Boehmer:2007ket,Chiou:2008nm}, and for comprehensive reviews, please refer to \cite{Perez:2017cmj,Zhang:2023yps}. Pioneering efforts also have been made to explore quantum gravity effects using EMRI systems \cite{Tu:2023xab,Liu:2024qci,Yang:2024lmj}. The authors in \cite{Liu:2024qci} points out that EMRIs provide more precise constraints compared to weak field experiments within the solar system. While in Refs. \cite{Yang:2024lmj,Tu:2023xab}, the authors employ the numerical kludge (NK) method to obtain gravitational waveforms from EMRIs on periodic orbits around an LQG-corrected BH proposed in \cite{Lewandowski:2022zce}. Their findings suggest that GW signals from EMRIs have the potential to probe quantum gravitational effects.

Recently, a covariant LQG-BH model has been proposed in \cite{Alonso-Bardaji:2021yls, Alonso-Bardaji:2022ear}.
A notable advancement in this model is the closure of the modified constraint algebra. This ensures the system provides a consistent, covariant, and unambiguous geometric representation, independent of the gauge choice on the phase space. The authors have subsequently extended this LQG-BH solution to incorporate charge within a cosmological context \cite{Alonso-Bardaji:2023niu} and have further investigated its coupling with matter \cite{Alonso-Bardaji:2021tvy, Alonso-Bardaji:2023vtl}. Additionally, numerous studies have explored various aspects of this model. For example, the quasinormal modes (QNMs) associated with this LQG-BH have been analyzed in \cite{Fu:2023drp, Moreira:2023cxy, Bolokhov:2023bwm}; the model’s potential extension to the Planck scale, along with considerations of a remnant, has been investigated in \cite{Sobrinho:2022zrp, Borges:2023fub}; and its gravitational lensing and optical characteristics have been analyzed in \cite{Soares:2023uup, Junior:2023xgl, Balali:2023ccr}. 

In this paper, we will study an EMRI, where a stellar-mass object spirals into a SMBH that incorporates covariant LQG corrections as described in \cite{Alonso-Bardaji:2021yls, Alonso-Bardaji:2022ear}. The most distinctive characteristic of this model is that the $tt$ component of the metric is consistent with the SS-BH, while quantum effect is only reflected through $r_0$ in the $rr$ component. We analyze the influence of the LQG effect on the orbital dynamics and waveform, and our results show that the $rr$ component of the metric contributes only at a subleading order to the energy and angular momentum fluxes.
For the extensive duration of EMRIs system, the accumulated dephasing of gravitational wave signals caused by the subleading order fluxes becomes detectable, allowing the corresponding quantum effects to be tested experimentally.
Then, we consider the detection capability for the LISA with faithfullness and Fisher information matrix (FIM).   

The overall structure of this paper is as follows: In section \ref{sec-1}, we introduce the LQG background and the derivation of the orbital evolution equation. Section \ref{sec-2} discusses gravitational radiation and derives the energy and angular momentum fluxes, incorporating corrections from loop quantum effects. Furthermore, we calculate the orbital evolution within the LQG background and analyze the effect of the LQG parameter $r_0$. In section \ref{sec-2}, we use the augmented analytic kludge (AAK) method to generate the GW signal and evaluate the error in detecting LQG parameter using the FIM. The conclusions are presented in section \ref{conclusion}.

\section{Quantum Schwarzschild spacetime and timelike geodesics}\label{sec-1}
\begin{figure}[H]
	\center{
		\includegraphics[scale=0.6]{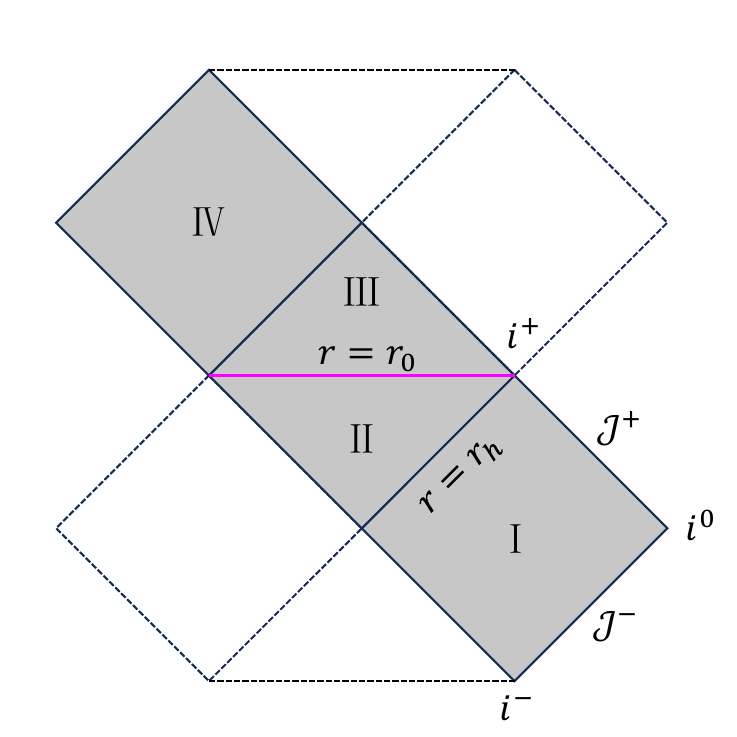}
		\caption{Penrose diagram of the quantum-corrected spacetime \eqref{metric}. The space-like hypersurface is located at the $r_0$, and the regions II and III correspond to the interior region of the black hole and white hole, respectively.}
		\label{Penrose}
	}
\end{figure}

In this work, we focus on a covariant LQG-BH with holonomy corrections proposed in \cite{Alonso-Bardaji:2021yls, Alonso-Bardaji:2022ear}
\begin{eqnarray} \label{metric}
&
ds^2=-f(r)dt^2+g(r)^{-1}dr^2+r^2d\Omega^2\,,
\nonumber
\\
&
f(r)=1-2M/r\,,\,\,\,\,\,\,\, g(r)=(1-r_0 M/r)f(r) \,.
\end{eqnarray}
where the horizon is given by $f(r_h)=0$ which is the same as that in the Schwarzschild spacetime.
The key difference between the metric \eqref{metric} and the Schwarzschild spacetime is the presence of a minimal space-like hypersurface $r_0$, which separates the trapped BH interior region from the anti-trapped white hole region (see the Fig. \ref{Penrose}) \cite{Alonso-Bardaji:2021yls, Alonso-Bardaji:2022ear}.
The LQG parameter $r_0$ is a dimensionless quantity and represents the degree of the deviation from the Schwarzschild BH. 
When $r_0=0$, the metric reduces to the Schwarzschild black hole. 

Considering a massive test particle moves on the equatorial plane $\theta=\pi/2$, the Lagrangian describing its motion around the background $\eqref{metric}$ is given by \cite{Chandrasekhar}
\begin{eqnarray}\label{lag}
	\mathcal{L}&=&\frac{1}{2} m g_{\mu \nu} \dot{x}^{\mu} \dot{x}^{\nu} \nonumber\\
	&=&\frac{1}{2} m \left(-f(r)\dot{t}^2+g(r)^{-1}\dot{r}^2+r^2\dot{\phi}^2\right)\,,
\end{eqnarray}
where $m$ is the mass of the test particle and the dot denotes the derivative with respect to the proper time $\tau$. For a static, spherically-symmetric system, the metric \eqref{metric} possesses two Killing vectors, $\xi^{a} = \left({\partial}/{\partial t}\right)^{a}$ and $\eta^{a} = \left({\partial}/{\partial \phi}\right)^{a}$, corresponding to the praticle's energy ${E}$ and angular momentum $L_z$, respectively.
The canonical momentum can be derived from the Lagrangian \eqref{lag} as follows:
\begin{eqnarray}\label{pt}
	&&p_{t} = - m f(r) \dot{t} =-E, \label{equ:II.B-(8)} \\[3mm]
	&&p_{r} = m \frac{1}{g\left(r\right)} \dot{r}, \\[3mm]
	&&p_{\phi} = m r^{2} \thinspace \dot{\phi}=L_z . \label{equ:II.B-(11)}
\end{eqnarray}
Applying the normalization condition $g_{\mu\nu}u^\mu u^\nu=-1$, the equation of motion reads as 
\begin{eqnarray}
\label{eom1}
\dot{t}&=&\frac{E}{m f(r)}, \\[3mm]
\label{eom2}
\dot{r}^2&=&g(r)\left(\frac{E^2}{m^2f(r)}-\frac{L^2}{m^2r^2}-1\right), \\[3mm]
\label{eom3}
\dot{\phi}&=&\frac{L_z}{m r^2}.
\end{eqnarray}
For the eccentric motion, we introduce the semi-latus rectum $p$ and the eccentricity $e$, and parameterize the radial coordinate by $\chi$ as 
\begin{eqnarray}\label{eccentric_orbit}
r(\chi)=\frac{Mp}{1+e \cos(\chi)}.
\end{eqnarray}
Recalling the Eq. \eqref{eom2} and combining with the condition for bound orbits $dr/d\tau=0$ at the two turning point $r_a=Mp/(1-e)$ and $r_p=Mp/(1+e)$, we have
\begin{eqnarray}
	\label{energy}
	E^2/m^2&=&\frac{(p-2-2e)(p-2+2e)}{p(p-3-e^2)}, \\
	\label{momentum}
	L_z^2/m^2&=&\frac{M^2p^2}{p-3-e^2},
\end{eqnarray}
where the bound orbits satisfy the inequalities $0\leq e < 1$ and $p>6+2e$.   Meanwhile, it is easy to see that the orbit energy $E$ and the angular momentum $L$ in the LQG background \eqref{metric} are consistent with the Schwarzschild case \cite{Cutler:1994pb, Hopper:2015icj}. 

For the eccentric orbits on the equatorial plane, there are two fundamental frequencies $\Omega_r$ and $\Omega_\phi$, which are related to the radial and azimuthal components, respectively:
\begin{eqnarray}
&&
\Omega_r=\frac{2\pi}{T_r}, \ \ \ T_r=\int_0^{2\pi}\frac{dt}{d\chi}d\chi\,,
\label{Omega_r}
\
\\
&&
\Omega_\phi=\frac{\Delta \phi}{T_r},  \ \ \ \ \Delta\phi=\int_0^{2\pi}\frac{d\phi}{d\chi}d\chi\,,
\label{Omega_phi}
\end{eqnarray}
where $T_r$ is the radial period, and $\Delta\phi$ represents the change in the azimuthal angle $\phi$. By substituting  Eqs. \eqref{eom1} and \eqref{eom3} into Eqs. \eqref{Omega_r} and \eqref{Omega_phi}, one can explicitly obtain  $\Omega_r$ and $\Omega_\phi$ in the large $p$ expansion as follows
\begin{eqnarray}
	\label{Omega_r_exp}
 \Omega_r&=&\frac{(1-e^2)^{3/2}}{M}p^{-3/2}-\frac{(1-e^2)^{5/2}(6+r_0)}{2M}p^{-5/2}+\mathcal{O}(p^{-7/2}), \\
 	\label{Omega_phi_exp}
 \Omega_\phi&=&\frac{(1-e^2)^{3/2}}{M}p^{-3/2}+\frac{e^2(1-e^2)^{3/2}(6+r_0)}{2M}p^{-5/2}+\mathcal{O}(p^{-7/2}).
\end{eqnarray}

\section{Fluxes and orbital evolution} \label{sec-2}

The previous discussion does not involve gravitational radiation. We now focus on investigating the GW radiation in the context of the LQG-BH and its impact on the modifications to orbital evolution. In the weak field approximation, we adopt the quadrupole formula to calculate the average GW fluxes. Using the relationship between the $\{E,L\}$ and the $\{p,e\}$, one can determine the orbital evolution of the inspiralling object \cite{Flanagan:2007tv,Thorne:1980ru,Ryan:1995zm,Maggiore,Liu:2024qci}. In this way, the quadrupole moment $Q^{ij}$ is given by 
\begin{eqnarray}\label{quardrupole}
Q^{ij}=\mathcal{M}^{ij}-\frac{1}{3}\delta^{ij} \mathcal{M}_{kk}\,,
\end{eqnarray}
where the mass moment of the test particle is 
\begin{eqnarray}\label{mass moment}
\mathcal{M}^{ij}=\mu x^ix^j\,,
\end{eqnarray}
with $\mu=m M/(m+M)$ as the reduced mass, which approximates to $\mu\simeq m$ in EMRIs.
The Cartesian coordinates $x^i$ are defined in terms of the spherical coordinate $\{r, \phi\}$ as $x^i=\{r \cos(\phi),r \sin(\phi),0\}$. In the weak field approximation, the average energy and angular momentum fluxes are given by
\begin{eqnarray}
	\label{dEdt}
&&\left<\frac{dE}{dt}\right>=\frac{(1-e^2)^{3/2}(96+292e^2+37e^4)m^2}{15M^2p^5} \nonumber\\
&&+\frac{e^2(1-e^2)^{3/2}(3(176+450e^2+53e^4)+(564+843e^2+70e^4)r_0)\mu^2}{15M^2p^6}+\mathcal{O}(p^{-7})\,,\\
\label{dLdt}
&&\left<\frac{dL_z}{dt}\right>=\frac{4(1-e^2)^{3/2}(8+7e^2)m^2}{5Mp^{7/2}}\nonumber \\
&&+\frac{e^2(1-e^2)^{3/2}(4(38+27e^2)+(104+41e^2)r_0)m^2}{5Mp^{9/2}}+\mathcal{O}(p^{-11/2})\,.
\end{eqnarray}
The LQG parameter $r_0$ presents only in the sub-leading order corrections to the fluxes, indicating that its influence becomes significant only in this order. 

Once the energy and angular fluxes are in hand, one can calculate the orbital evolution with the gravitational radiation reaction.
Based on the adiabatic approximation, we assume that the reduction in orbital energy and angular momentum is fully converted into the averaged radiated energy and angular momentum fluxes 
\begin{eqnarray}
\left<\frac{dE}{dt}\right>_{GW}=-\left<\frac{dE}{dt}\right>=-\mu \dot{E},\ \ \
\left<\frac{dL_z}{dt}\right>_{GW}=-\left<\frac{dL_z}{dt}\right>= -\mu \dot{L_z}.
\end{eqnarray}
Therefore, the $\dot{E}$ and $\dot{L_z}$ can be rewritten in terms of the $p$ and $e$ as follows:
\begin{eqnarray}
-\dot{E}=m \frac{\partial E}{\partial p}\frac{dp}{dt}+m\frac{\partial E}{\partial e}\frac{de}{dt}, \\
-\dot{L_z}=m \frac{\partial L_z}{\partial p}\frac{dp}{dt}+m\frac{\partial L_z}{\partial e}\frac{de}{dt}.
\end{eqnarray}
Using Eqs.\eqref{energy} and \eqref{momentum}, we have
\begin{eqnarray}
\label{dpdt}
\frac{dp}{dt}&=&\frac{2(p-3-e^2)^{1/2}}{(p-6-2e)(p-6+2e)}\bigg[p^{3/2}(p-2-2e)^{1/2}(p-2+2e)^{1/2}\dot{E} \nonumber \\
&-&(p-4)^2\dot{L_z}/M\bigg] ,\\
\label{dedt}
\frac{de}{dt}&=&\frac{(p-3-e^2)^{1/2}}{ep(p-6-2e)(p-6+2e)}\bigg[-p^{3/2}(p-6-2e^2)(p-2-2e)^{1/2}(p-2+2e)^{1/2}\dot{E} \nonumber \\
&+&(1-e^2)((p-2)(p-6)+4e^2)\dot{L}/M\bigg]. 
\end{eqnarray}
From the above equations, it is evident that the orbital evolution parameters $\{dp/dt, de/dt\}$ depend on the LQG parameter $r_0$ through the energy fluxes \eqref{dEdt} and angular momentum fluxes \eqref{dLdt}.
Without a doubt, the effects of LQG on the orbital evolution parameters $\{dp/dt, de/dt\}$ become significant only at the subleading order.

\begin{figure}[H]
	\center{
		\includegraphics[scale=0.85]{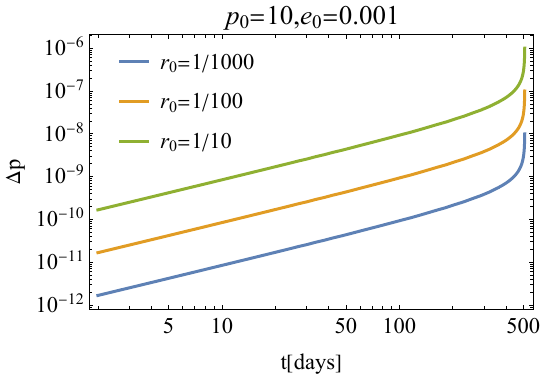}\hspace{0.4cm}
		\includegraphics[scale=0.85]{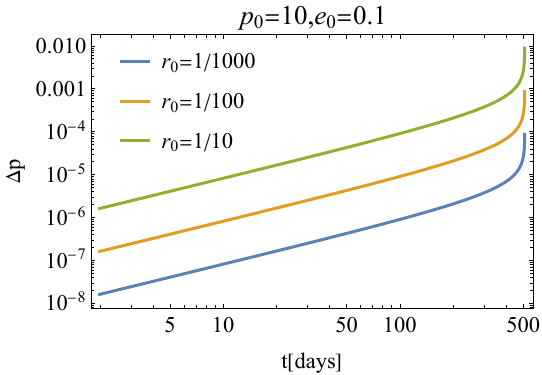}
		\caption{The difference of the semi-latus $\Delta p=p(\text{SS})-p(\text{LQG})$ with different LQG parameter $r_0$. The left and right plots denote the initial condition $(p_0=10, e_0=0.001)$ and  $(p_0=10, e_0=0.1)$, respectively. }
		\label{Orbital_p}
	}
\end{figure}

\begin{figure}[H]
	\center{
		\includegraphics[scale=0.85]{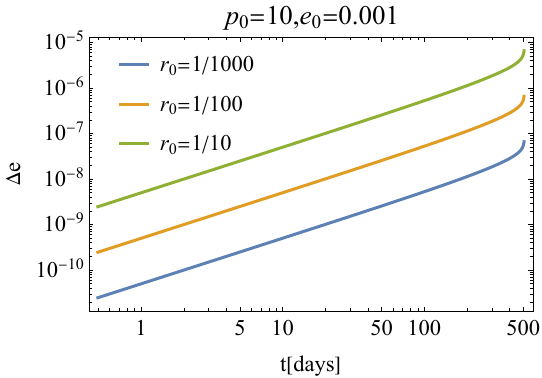}\hspace{0.4cm}
		\includegraphics[scale=0.85]{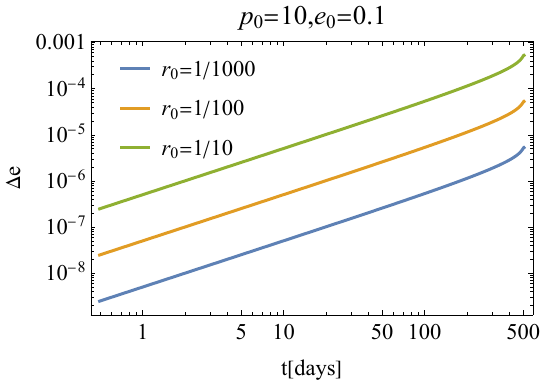}
		\caption{The difference of the eccentricity $\Delta e=e(\text{SS})-e(\text{LQG})$ with different LQG parameter $r_0$. The left and right plots denotes the initial condition $(p_0=10, e_0=0.001)$ and  $(p_0=10, e_0=0.1)$, respectively.}
		\label{Orbital_e}
	}
\end{figure}

Since the effects of LQG appear only at the subleading order, it is more illustrative to show the relative changes, $\Delta p(t)$ and $\Delta e(t)$, in the orbital evolution parameters $\{dp/dt, de/dt\}$ by subtracting the Schwarzschild contribution.
Figures \ref{Orbital_p} and \ref{Orbital_e} display the relative changes, $\Delta p(t)$ and $\Delta e(t)$, after subtracting the Schwarzschild contribution, for different values of the effective quantum parameter $r_0$. 
It can be observed that the relative changes in the $\{\Delta p(t), \Delta e(t)\}$ increase over time. Notably, the magnitude of these changes becomes markedly more pronounced, underscoring the increasingly significant influence of quantum effects on orbital dynamics. 


\begin{figure}[H]
	\center{
		\includegraphics[scale=0.85]{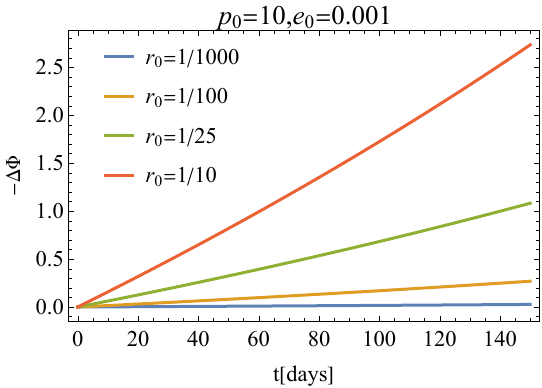}\hspace{0.4cm}
		\includegraphics[scale=0.85]{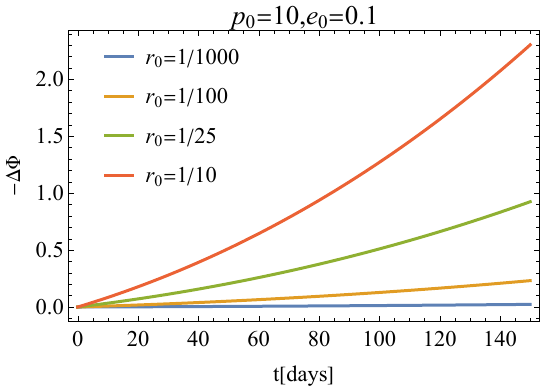}
		\caption{The accumulated orbital phase difference $\Delta \Phi=\Phi_\phi(\text{SS})-\Phi_\phi(\text{LQG})$ with different LQG parameter $r_0$. The left and right plots denotes the initial condition $(p_0=10, e_0=0.001)$ and  $(p_0=10, e_0=0.1)$, respectively.}
		\label{Orbital_Phi}
	}
\end{figure}

In the eccentric orbit, there are the two fundamental frequencies $\{\Omega_\phi, \Omega_r\}$, corresponding to two phases, $\{\Phi_\phi, \Phi_r\}$. The average rate change of phases is given by
\begin{eqnarray}\label{phases}
\frac{d\Phi_i}{dt}=\left<\Omega_i(p(t),e(t))\right>=\frac{1}{T_r}\int_{0}^{2\pi}\Omega_i(p(t),e(t)) \frac{dt}{d\chi}d\chi\,.
\end{eqnarray}
Recalling Eqs.\eqref{Omega_r_exp}, \eqref{Omega_phi_exp}, \eqref{dpdt}, and \eqref{dedt}, the radial and azimuthal phases $\Phi_i (i=t,r)$ can be numerically solved with the initial condition $\Phi_i(0)=0$. 
Fig.\ref{Orbital_Phi} shows the accumulated dephasing $\Delta\Phi=\Phi_\phi(\text{SS})-\Phi_\phi(\text{LQG})$ as a function of time for different LQG parameters $r_0$. It can be observed that the dephasing $\Delta\Phi$ increases with rising $r_0$. This indicates that, as time accumulates, the phase differences manifest in the gravitational waveforms. In the next section, we shall analyze the deviations in gravitational waveforms from those of a Schwarzschild black hole for different LQG parameters $r_0$.

\section{waveform and parameter estimation}\label{sec-3}

In this section, we employ the AAK approach to generate EMRI waveforms and perform the FIM estimation to constrain the LQG parameter.
Compared to the AK method, the AAK waveform improves the accuracy of waveform templates without significantly increasing computational time.  Ref.\cite{Chua:2017ujo} points out that the AAK waveform can detect at least an order of magnitude more events than the AK waveform across the 12 astrophysical EMRI-population models.
Additionally, the AAK waveform is physically consistent with the NK waveform. However, the AAK method has approximately achieved an order of magnitude increase in computational speed compared to the NK method \cite{Chua:2017ujo}.

For convenience, we adopt a detector-adapted frame, where the unit vector $\hat{\mathbf{r}}$ points from the detector towards the source, and the other unit vectors $\hat{\mathbf{p}}$ and $\hat{\mathbf{q}}$ can be defined as 
\begin{eqnarray}
\hat{\mathbf{p}}=\frac{\hat{\mathbf{r}} \times \hat{\mathbf{L}}}{|\hat{\mathbf{r}} \times \hat{\mathbf{L}}|}, \quad \hat{\mathbf{q}}=\hat{\mathbf{p}} \times \hat{\mathbf{r}}.
\end{eqnarray}
In this frame, the waveform, in the transverse-traceless gauge, can be described by the quadrupole approximation
\begin{eqnarray}\label{hij}
h_{ij}=\frac{2}{D_L}(P_{ik}P_{jl}-\frac{1}{2}P_{ij}P_{kl})\ddot{I}^{kl},\ \ \  h^{\{+,\times\}} =\frac{1}{2}h^{ij}H_{ij}^{\{+,\times\}}
\end{eqnarray}
where $D_L$ is the source luminosity distance, the projection tensors $P_{ij}$ and the polarization basis tensor $H_{ij}^{\{+,\times\}}$ are defined by 
\begin{eqnarray}
H_{i j}^{+}=\hat{p}_{i} \hat{p}_{j}-\hat{q}_{i} \hat{q}_{j}, \quad H_{i j}^{\times}=\hat{p}_{i} \hat{q}_{j}+\hat{q}_{i} \hat{p}_{j}, \quad P_{i j}=\delta_{i j}-\hat{r}_{i} \hat{r}_{j}.
\end{eqnarray}
By the decomposition of $n$-harmonic components, the two polarization waveform can be expressed as \cite{Barack:2003fp}
\begin{eqnarray}
\begin{array}{l}
h^{\{+, \times\}}=\sum_{n} A_{n}^{\{+, \times\}}, \\
A_{n}^{+}=\left[1+(\hat{\mathbf{r}} \cdot \hat{\mathbf{L}})^{2}\right]\left[b_{n} \sin (2 \gamma)-a_{n} \cos (2 \gamma)\right]+\left[1-(\hat{\mathbf{r}} \cdot \hat{\mathbf{L}})^{2}\right] c_{n}, \\
A_{n}^{\times}=2(\hat{\mathbf{r}} \cdot \hat{\mathbf{L}})\left[b_{n} \cos (2 \gamma)+a_{n} \sin (2 \gamma)\right] .
\end{array}
\end{eqnarray}
Here, $\gamma=\Phi_\phi-\Phi_r$ represents the direction of the pericenter relative to the unit vector $\hat{\mathbf{r}}$.
The coefficients $(a_n, b_n, c_n)$ can be expressed by the Bessel function of the first kind $J_n$ as    
\begin{eqnarray}
a_{n}&= & -n \mathcal{A}\left[J_{n-2}(n e)-2 e J_{n-1}(n e)+(2 / n) J_{n}(n e)\right. \nonumber\\
&& \left.+2 e J_{n+1}(n e)-J_{n+2}(n e)\right] \cos \left[n \Phi_{r}(t)\right], \\
b_{n}&= & -n \mathcal{A}\left(1-e^{2}\right)^{1 / 2}\left[J_{n-2}(n e)-2 J_{n}(n e)\right. \nonumber\\
&& \left.+J_{n+2}(n e)\right] \sin \left[n \Phi_{r}(t)\right], \\
c_{n}&= & 2 \mathcal{A} J_{n}(n e) \cos \left[n \Phi_{r}(t)\right], 
\end{eqnarray}
with
\begin{eqnarray}
\mathcal{A}=(M\Omega_\phi)^{2/3}m/D_L.
\end{eqnarray}

\begin{figure}[H]
	\center{
		\subfigure[$r_0=1/1000$]{\includegraphics[scale=1.2]{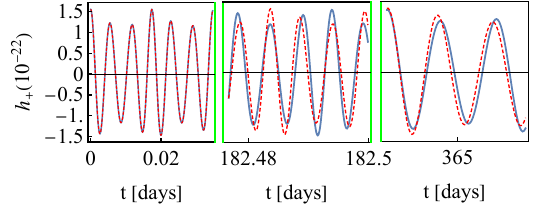}}
		\subfigure[ $r_0=1/100$]{\includegraphics[scale=1.2]{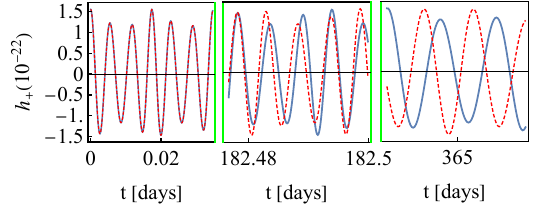}}
		\subfigure[ $r_0=1/10$]{\includegraphics[scale=1.2]{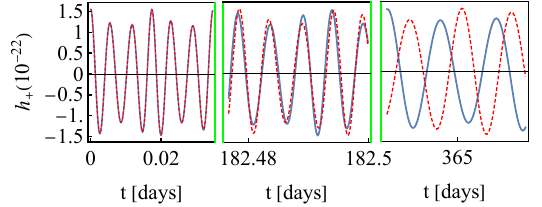}}
		\caption{The waveform with different LQG parameter values of $r_0$ for the initial eccentricity $e_0=0.1$ and semi-latus rectum $p_0$=10. The blue line represents the Schwarzschild case, while the red dashed line denotes the LQG-BH case.}
		\label{Waveform}
	}
\end{figure}
Fig.\ref{Waveform} presents the gravitational waveforms corresponding to various values of $r_0$, given an initial position $p_0 = 10$ and $e_0 = 0.1$.
Without loss of generality, we set the mass of SMBH to $M=10^6 M_\odot$ and the mass of the stellar-mass BH to $m=10 M_\odot$.
The waveforms reveal that for large values of $r_0$, there is a significant deviation in the waveform’s phase, as illustrated in the second and third plots of Fig.\ref{Waveform}. This pronounced phase difference suggests that the EMRI system could serve as a sensitive probe to detect the imprints of LQG. Conversely, as shown in Fig.\ref{Waveform}, when $r_0=1/1000$ the influence of LQG are extremely weak, resulting in waveforms that remain nearly indistinguishable even after prolonged evolution.
To further quantify the detection capability of LQG effects, we perform a faithfulness between the Schwarzschild BH and the LQG-BH, providing a potential pathway to probe quantum gravitational effects in astronomical observations

The LISA detector has three arms, which can be seen as two Michelson interferometers, labeled I and II.
The GW strains can be  described by the same harmonic decomposition as \cite{Cutler:1997ta}
\begin{eqnarray}
h_{\text{I,II}}(t)=h_+(t)F_{\text{I,II}}^++h_\times(t)F_{\text{I,II}}^{\times}
\end{eqnarray}
where the $F_{\text{I,II}}^{+,\times}$ are the antenna pattern functions, and it only depends on the source orientation $(\theta_s, \phi_s)$ and the orbital angular direction $(\theta_1, \phi_1)$ \cite{Barack:2003fp,Cutler:1997ta}.

\begin{figure}[H]
	\center{
		\includegraphics[scale=0.9]{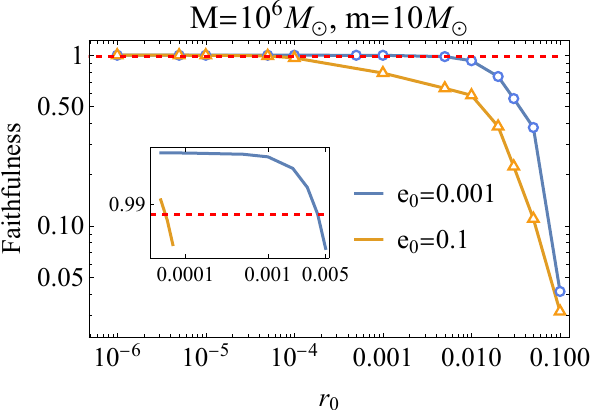}
		\caption{The faithfulness of the GW signal between the Schwarzchild BH and LQG-BH for different $e_0$. The blue and orange lines correspond to the $e_0=0.001$ and $e_0=0.1$, respectively. }
		\label{Faithfulness}
	}
\end{figure}

The signal-to-noise ratio (SNR) helps us understand the observability of a signal, thereby allowing us to evaluate whether a detector can effectively detect EMRI events. By employing the noise-weighted inner product, one can compute the SNR of the waveform $h$ as 
\begin{eqnarray}
\rho: =\sqrt{\left\langle h\mid h\right \rangle}=2\left[\int_{0}^{\infty}\frac{\Tilde{h}(f)\Tilde{h}^*(f)}{S_n(f)} df\right]^{1/2}\,,
\end{eqnarray}
where $S_n$ is the power spectral density (PSD) for the LISA detector. Furthermore, we can define the faithfulness between the two signals by their cross-correlation as
\begin{eqnarray}\label{faithfulness}
\mathcal{F}\left[h_{1}, h_{2}\right]=\max _{\left\{t_{c}, \phi_{c}\right\}} \frac{\left\langle h_{1} \mid h_{2}\right\rangle}{\sqrt{\left\langle h_{1} \mid h_{1}\right\rangle\left\langle h_{2} \mid h_{2}\right\rangle}}\,.
\end{eqnarray}
Here, $(t_c, \phi_c)$ represent the time and phase offsets, respectively \cite{Lindblom:2008cm}. The faithfulness, as defined in \eqref{faithfulness}, quantifies the distinguishability between two GW signals. When $\mathcal{F} = 1$, the two signals are identical, indicating perfect overlap. Conversely, $\mathcal{F} = 0$ corresponds to two completely orthogonal waveforms, indicating no similarity. As argued in Ref.\cite{Chatziioannou:2017tdw}, LISA can effectively distinguish between two signals if $\mathcal{F}\leq 0.988$. We adopt this threshold as the criterion for determining signal mismatch. Fig.\ref{Faithfulness} shows faithfulness plotted as a function of $r_0$ for different values of $e_0$. It was observed that, through one-year observation, the faithfulness becomes worse as $r_0$ increases. It suggests that EMRIs may provide a viable method for detecting the effects of quantum gravity. Specially, for the initial eccentricity $e_0=0.001$, LISA is capable of distinguishing the LQG signal in the region where $r_0\geq 0.00403$ (as indicated by the blue line in Fig.\ref{Faithfulness}). However, for larger values of $e_0$, the required $r_0$ for distinguishability decreases, meaning that LISA can detect the quantum gravity effects even at smaller values of $r_0$ . For instance, when $e_0=0.1$, the LQG effect can be detected by LISA around $r_0 \geq 0.000057$, as shown by the orange line in Fig.\ref{Faithfulness}.

Since faithfulness does not account for the correlations between the parameters, this may lead to imprecision on the constraint of the $r_0$. To more accurately estimate the LQG parameter, we execute the Fisher method to estimate the parameters. In the large SNR limit, we assume that the best-fit parameter satisfies the Gaussian distribution, and the Fisher information matrix $\Gamma_{ij}$ is given by
\begin{eqnarray}
\label{FIMatrix}
\Gamma_{i j}=\left\langle\left.\frac{\partial h}{\partial \xi_{i}}\right| \frac{\partial h}{\partial \xi_{j}}\right\rangle_{\xi=\hat{\xi}}\,.
\end{eqnarray}
Here, $\xi$ denotes the phase space, which can be described by ten parameters as 
\begin{eqnarray}
\xi=(\ln{M}, \ln{m}, p_0, e_0, r_0, \theta_s, \phi_s, \theta_1, \phi_1, D_L).
\end{eqnarray}
From Eq.\eqref{FIMatrix}, one can obtain the variance-covariance matrix by inverting the FIM, i.e.
\begin{eqnarray}
\label{vc_matrix}
\Sigma_{ij}\equiv \left\langle\delta\xi_i \delta\xi_j\right\rangle=(\Gamma^{-1})_{ij}.
\end{eqnarray}
The statistical error on $\xi$ are provided by the diagonal element of \blue{Eq.} \eqref{vc_matrix}
\begin{eqnarray}
\sigma_i=\Sigma_{ii}^{1/2}.
\end{eqnarray}
Since the space-based GW detector has the triangle configuration, which forms a network of two L-shaped detectors. The total SNR and the total covariance matrix is the sum of the two L-shaped detectors $h_1$ and $h_2$, which can be expressed as
\begin{eqnarray}
\rho&=&\sqrt{\rho_1^2+\rho_2^2}=\sqrt{\left\langle h_1\mid h_1\right\rangle+\left\langle h_2\mid h_2\right\rangle}, \\
\sigma_i^2&=&(\Gamma_1+\Gamma_2)_{ii}^{-1}.
\end{eqnarray}

\begin{figure}[H]
	\center{
		\includegraphics[scale=0.5]{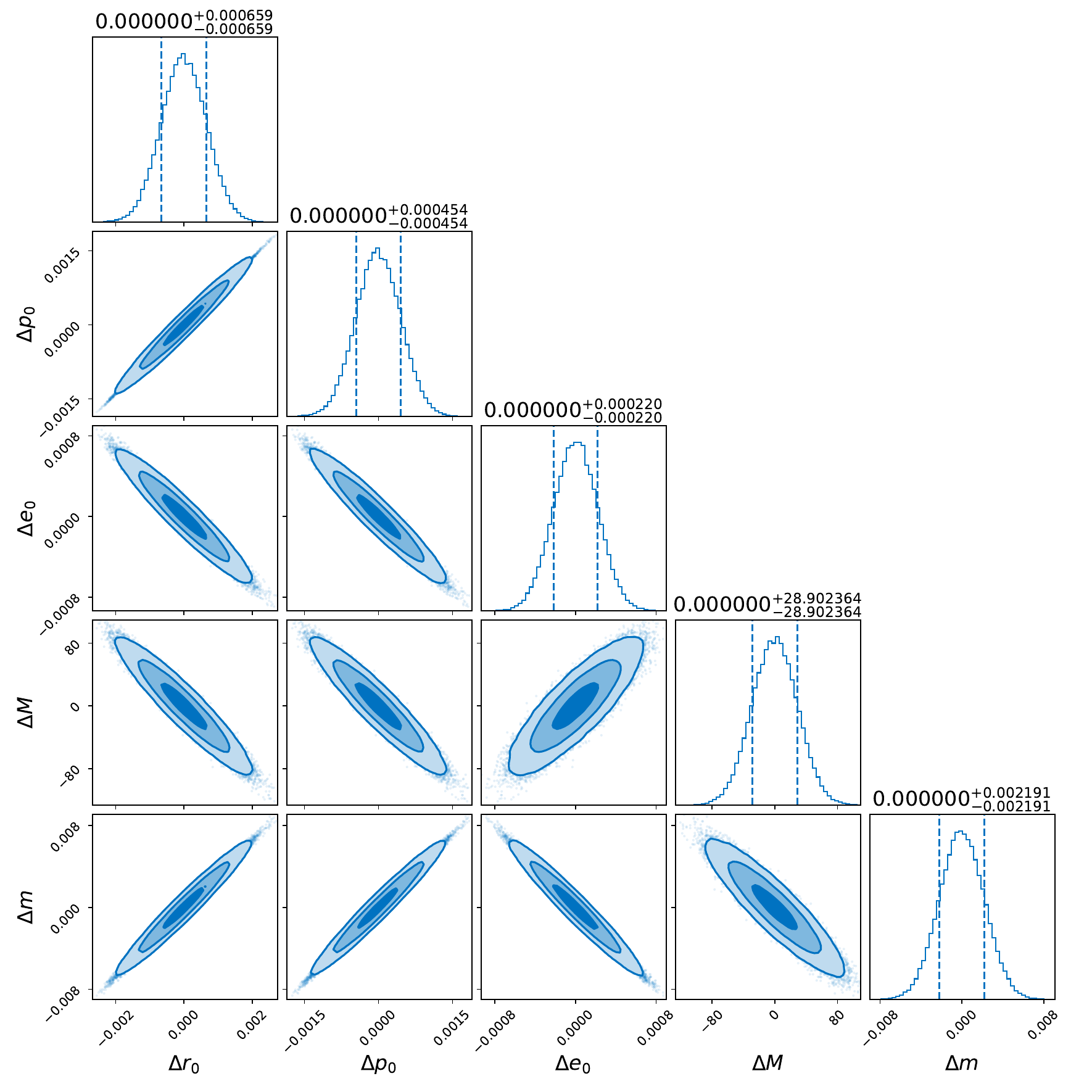}
		\caption{Corner plot for the probability distribution of intrinsic parameters for LISA. Diagonal boxes refer to marginalized distributions. Vertical lines show the 1-$\sigma$ interval for each waveform parameter. The contours correspond to the $68\%$, $95\%$, and $99\%$ probability confidence intervals.}
		\label{FIM}
	}
\end{figure}
We fix the source angles at $\theta_s=\pi/3,\ \phi_s=\pi/2,\ \theta_1=\pi/4,\ \phi_1=\pi/4$, and choose the system parameters as $M=10^6 M_\odot$, $m=10 M_\odot$, $D_L=1$ Gpc, $r_0=0$, $e_0=0.1$ and $p_0=9.47446$, ensuring that evolution period is one year before the final plunge.  Fig.\ref{FIM} depicted the probability distributions for the binary masses, initial orbit, and the LQG parameter $r_0$. The results suggest that after one-year observation, the error in the parameter $r_0$ can be constrained to approximately $\Delta r_0=6.59\times 10^{-4}$ with SNR=49.

\section{Conclusion} \label{conclusion}

EMRI systems are among the most significant GW sources for future space-based GW detectors, such as LISA, TianQin, and Taiji. The GWs emitted by these systems carry rich and detailed information about the spacetime geometry surrounding BHs. Analyzing EMRI systems offers an unique opportunity to probe the intricate properties of BHs, explore the fundamental nature of spacetime, and deepen our understanding of GW physics. Furthermore, these systems serve as a natural laboratory for testing or challenging existing theories of gravity, including GR and alternative gravity theories. They may also provide insights into new physical phenomena, such as potential quantum gravity effects, thereby opening new frontiers in astrophysics and fundamental physics.

In this paper, we investigate quantum gravity effects in EMRI systems by modeling the SMBH with a covariant LQG framework \cite{Alonso-Bardaji:2021yls,Alonso-Bardaji:2022ear}. We focus on calculating the average GW fluxes and find that, the LQG parameter affects only the subleading order corrections to the average energy and angular momentum fluxes.
Despite the initial appearance of quantum gravity effects at subleading orders, their influence on EMRI systems becomes increasingly significant over prolonged periods of evolution. EMRI systems, which produce GW signals that can last for years or even decades, are particularly well-suited for long-term observational studies. As these systems evolve, quantum gravity effects progressively alter orbital dynamics and waveforms, leading to detectable deviations. 

Furthermore, we also apply faithfulness to constrain LQG parameters. Our findings underscore the potential of EMRI systems to identify LQG signatures. Notably, our results indicate that the LISA detector can distinguish LQG signals when the quantum parameter $r_0\geq 0.00403$ with an initial eccentricity of $e_0=0.001$. For higher initial eccentricities, smaller values of $r_0$ is sufficient for the signal to be distinguished. 
Specifically, for the $e_0=0.1$, the detection limit for the LQG effect is $r_0\geq0.000057$.
Additionally, using the FIM method, we estimate that after one year of observation, the uncertainty in $r_0$ can be reduced to approximately $\Delta r_0=6.59\times 10^{-4}$ with a SNR of $49$. These findings highlight the potential of EMRI systems as a powerful tool for detecting LQG effects and advancing our understanding of quantum gravity through gravitational wave astronomy.

\acknowledgments

We are very grateful to Yunlong Liu for the valuable discussions. This work is supported by National Key R$\&$D Program of China (No. 2020YFC2201400), the Natural Science Foundation of China under Grants Nos. 12375055, 12347159 and 12375056.

\appendix

\bibliographystyle{style1}
\bibliography{Ref}
\end{document}